\documentstyle[psfig]{l-aa}     
\def\grs{{GRS~1739--278}}
\def\gra{{\sl Granat}}
\def\ros{{\sl ROSAT}}
\def\etal{{et\,al. }}

\def\msun{M$_{\odot}$}

\def\grad{$^\circ$}
\newbox\grsign \setbox\grsign=\hbox{$>$}
\newdimen\grdimen \grdimen=\ht\grsign
\newbox\laxbox \newbox\gaxbox
\setbox\gaxbox=\hbox{\raise.5ex\hbox{$>$}\llap
     {\lower.5ex\hbox{$\sim$}}}\ht1=\grdimen\dp1=0pt
\setbox\laxbox=\hbox{\raise.5ex\hbox{$<$}\llap
     {\lower.5ex\hbox{$\sim$}}}\ht2=\grdimen\dp2=0pt
\def\gax{\mathrel{\copy\gaxbox}}

\def\degs{\ifmmode ^{\circ}\else$^{\circ}$\fi}
\def\amin{\ifmmode ^{\prime}\else$^{\prime}$\fi}
\def\asec{\ifmmode ^{\prime\prime}\else$^{\prime\prime}$\fi}
\def\fss{\hbox{$.\!\!^{\rm s}$}}        
\def\fdg{\hbox{$.\!\!^\circ$}}          
\def\h{$^{\rm h}$}\def\m{$^{\rm m}$}

\begin{document}
 
   \thesaurus{06         
              (02.01.2;  
               08.02.3;  
               13.25.5)  
             }

   \title{ROSAT observation of GRS 1739--278}

   \author{J. Greiner, K. Dennerl, P. Predehl}

   \offprints{Greiner, jcg@mpe-garching.mpg.de}
 
  \institute{Max-Planck-Institut f\"ur extraterrestrische Physik,
             85740 Garching, Germany}

   \date{Received July 11, 1996; accepted August 30, 1996}
 
   \maketitle

   \begin{abstract}
We have observed \grs\, with the \ros\, HRI on March 31, 1996.
The improved X-ray position proves the identification with an earlier 
discovered variable radio source. We 
derive an extinction of $A_{\rm V} = 14\pm2$ from the dust scattering halo,
which implies a distance of d $\approx$ 6--8.5 kpc. 
Thus, \grs\, radiates at least near the Eddington limit for a 1 \msun\, compact
object.

The existence of the dust scattering halo necessitates a prolonged X-ray 
activity period prior to the ROSAT observation. Combined with the non-detection
by Granat during an earlier galactic center observation, we speculate that 
the X-ray turn-on of \grs\, happened in Nov./Dec. 1995.


   \end{abstract}
 
\section{Introduction}

A new X-ray transient (GRS~1739--278) was discovered by the SIGMA instrument
onboard \gra\, on March 18, 1996 (Paul \etal 1996) at a level of 80 mCrab in the
40--75 keV range. The source was detected also in an observation taken already 
on March 16.8 UT showing that the source turned on earlier.
\grs\, was also detected with the TTM experiment on the Mir-Kvant Space
Station as early as February 28 as a bright source (200 mCrab in the 2--27 keV
range) rising to 200 mCrab on March 1, 1996 (Borozdin \etal 1996). 

\grs\, was also observed with the Proportional Counter Array (PCA) on RXTE
on March 31.754--31.862 UT and found at 490 mCrab in the 2--60 keV range
(Takeshima \etal 1996).
Using slew data taken on other occasions, the following flux values were
also reported: 810 mCrab on March 16, 430  mCrab on March 22 and 300 mCrab on 
April 21. No oscillations were found with pe\-riods between 0.002 and 256 sec.

Subsequent radio observations at the VLA of the 1\amin\, radius TTM X-ray 
error box (Borozdin \etal 1996) revealed a variable radio source at 
R.A. = 17\h42\m40\fss03, Decl.=\mbox{--27\grad44\amin52\asec}\, as a probable 
counterpart (Hjellming \& Rupen 1996).

During the writing of this letter, the discovery of the optical and infrared 
counterpart of the radio source is reported at R = 20.5 mag, J = 16.0 mag 
and K = 14.7 mag (Mirabel \etal 1996).

Here, we report the results of a short \ros\, target of opportunity observation 
of \grs.

\section{Observational results}

\subsection{Position and Intensity}

\grs\, was observed with the \ros\, high resolution imager (HRI) on 
March 31.142, 1996 for a total of 620 sec, and was clearly detected at 1\amin\, 
off-axis angle.
The best fit position (2000.0) of \grs\, from this \ros\, HRI detection is:
R.A. = 17\h42\m40\fss3, Decl. = --27\grad44\amin54\asec\, with a systematic
boresight error of $\pm$8\asec\, and was reported already earlier 
(Dennerl \& Greiner 1996).
This centroid X-ray position is only 4\asec\, off the position of the 
variable radio source and thus suggests that the latter is indeed the 
radio counterpart of \grs.

Extracting all photons within 1600\asec\, (see below for the reason) yields,
after background subtraction and 
vignetting correction, a mean HRI count rate of 20 cts/sec. 

For the lightcurve reduction a much smaller extraction radius (10\asec) has 
been chosen in order to avoid the dust scattering halo which would smear out 
temporal variations.
Since photons at increasing distances from the central source have travelled
longer and thus are emitted earlier any lightcurve of the total detected
photons would include an averaging in time of the source intensity.
After background subtraction, vignetting and deadtime correction
the lightcurve shows no evidence for short-term variability in the 
1--200 sec range in excess of a Poisson distribution.

Adopting the spectral fit parameters for the low-energy range (2--15 keV)
of the near-simultaneous (14 hours after the
\ros\, observations) PCA/RXTE observation (i.e. a powerlaw model with
photon index of --4.1 at a normalisation of 0.51 photons/cm$^2$/s at 1 keV
which corresponds to 190 mCrab; Takeshima 1996), 
we have modelled the amount of column density in order to 
achieve the  measured total HRI countrate of 20 cts/s, resulting in
N$_{\rm H}$ = (2.6$\pm$0.5)$\times$10$^{22}$ cm$^{-2}$. This is somewhat 
below the value of (4.1$\pm$0.7)$\times$10$^{22}$ reported by Borozdin \etal 
(1996) for the
TTM spectral fit of a pure powerlaw model, but they also find a smaller
absorbing column for a two component model fit (without specifying the value).

\subsection{Dust scattering halo}

\begin{figure}[th]
      \centering{
      \hspace*{.001cm}
      \vbox{\psfig{figure=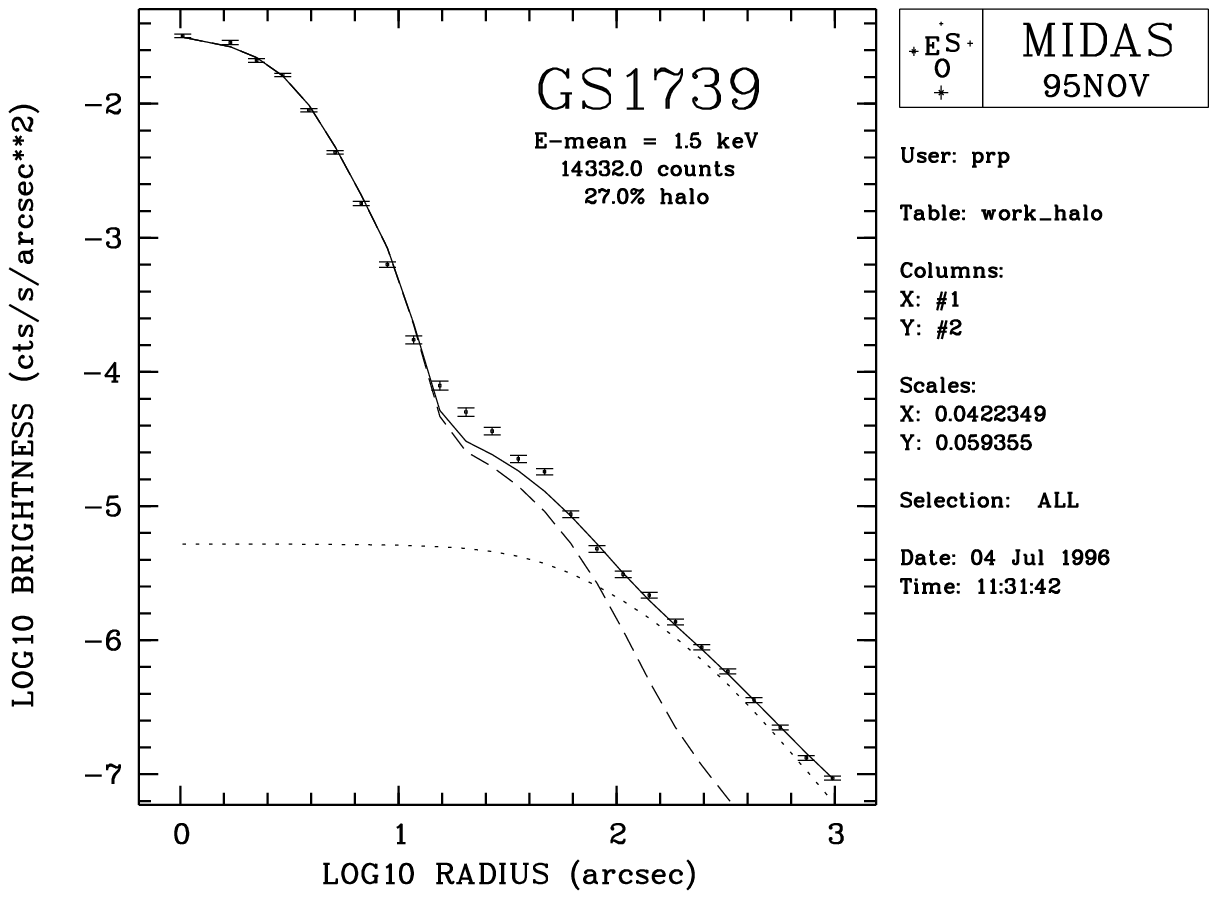,width=8.7cm,%
          bbllx=2.9cm,bblly=4.0cm,bburx=12.05cm,bbury=13.1cm,clip=}}\par}
\begin{picture}(12,25)\thicklines
\put(-30,100){\vector(0,2){15}}
\put(-60,90){dust scattering}
\put(55,65){\vector(2,0){15}}
\put(-40,63){ROSAT telescope PSF}
\end{picture}
     \vspace*{-1.cm}
      \caption[dust]{Radial profile of the X-ray emission of \grs. The dust
       scattering halo dominates the brightness at distances larger than
       about 100\asec\, and extends over the full HRI field of view.}
      \label{dust}
    \end{figure}

The observed radial profile of \grs\, in the HRI is considerably wider than 
the instrument point spread function. This is due to the scattering of the
X-rays of \grs\, by interstellar dust (galactic latitude of bII=1\fdg2).
Since this scattering
effect does not alter the number of detected photons, the extraction radius
for deducing the total mean count rate was chosen to be as large as
possible, i.e. 1600\asec\, in this case where there is no other source
in the field of view.
The  relative intensity of the halo
is 27\% at a mean energy E = 1.5 keV (Fig. \ref{dust}). 
The fractional halo intensity has been calculated by comparing the measured
radial surface brightness distribution with a HRI model point
response function and fitting the residual with a model halo
(Predehl \etal 1991), using a  common powerlaw grain size model
(Mathis, Rumpl and Nordsieck 1977) and an
uniform dust distribution between observer and source.
Since we have no spectral information for each photon in the HRI, we have
folded the model (see previous section) through the \ros\, mirror effective 
area and calculated the mean energy of all photons, as would have
been detected with the ROSAT HRI, to be 1.47 keV.

From observations of other X-ray sources having known optical counterparts,
correlations of the deduced effective optical depth
$\tau_{eff}$(1 keV) with the optical extinction 
(A$_V$/$\tau_{\rm eff}$(1 keV)$\approx$17) and with the hydrogen column
density (N$_{\rm H}$[10$^{21}$] cm$^{-2}$/$\tau_{\rm eff}$(1 keV)$\approx$23) 
have been deduced (Predehl \& Schmitt 1995). 
Applying these relations to \grs\, results in
$A_{\rm V}$ = 12$\pm$3 or $N_{\rm H}$=(1.6$\pm$0.4)$\times$10$^{22}$ cm$^{-2}$. 
Given the many uncertainties (not really contemporaneous \ros/RXTE
measurement, absolute flux calibration of the PCA/XTE, missing information
of the energy of each scattered photon in the HRI) this value of 
$N_{\rm H}$ is in 
reasonable agreement with the one derived in section 2.1. In addition, the
dust scattering halo intensity corresponds to rather different times
of emission from \grs\, due to the travel time delay (see below), and any 
variability
will result in varying fractional halo intensities. Inspecting the all-sky
monitor (ASM) lightcurve of the RXTE satellite shows that the \ros\, observation
occurred shortly after a major flare in the 2--10 keV band intensity,
and it is thus conceivable that the halo intensity is from a less intense
mean state of the source. Consequently, our above estimate of $A_{\rm V}$
is probably a lower limit. We therefore use
for the following $A_{\rm V} = 14\pm2$ and, accordingly,
$N_{\rm H}$=(2.0$\pm$0.3)$\times$10$^{22}$ cm$^{-2}$.

\subsection{Distance and luminosity}

Using a mean extinction along the galactic plane
of A$_{\rm V}$ = 1.9 mag/kpc (Allen 1973), the above A$_{\rm V}$ range implies
a distance of \grs\, of about 6--8.5 kpc. The A$_{\rm V}$ contribution
towards the galactic center along distance shows a rapid increase within the
first 1 kpc but then remains rather flat up to 5 kpc (Neckel \& Klare 1980),
suggesting the latter number to be a lower limit for the distance of \grs.

With the RXTE spectral parameters the unabsorbed flux is 
2.8$\times$10$^{-6}$ erg/cm$^2$/s corresponding to an unabsorbed 
luminosity of 1.6$\times$10$^{40}$ / (d/7\,kpc)$^2$ erg/s (0.1--2.4 keV).
This is an extraordinary high value for a galactic X-ray transient.
We note that the derivation of the flux and luminosity depends very sensitively 
on the assumption about the spectral shape at a few keV. For instance,
adopting a --3 photon index power law model with 
N$_{\rm H}$=2.0$\times$10$^{22}$ cm$^{-2}$ and adjusting the normalisation 
such that the HRI countrate is reproduced, reduces the unabsorbed flux by a 
factor of 10. We therefore caution that the --4.1 photon powerlaw fit of the
near-simultaneous PCA/RXTE data is an oversimplification, and any 
intrinsically curved spectral model (like disk blackbody or bremsstrahlung)
will reduce the X-ray luminosity drastically compared to the above value.
However, it is worth to realize that the estimates of A$_{\rm V}$ and the 
distance are much less affected than the luminosity estimate.

In order to derive an approximate lower limit for the soft X-ray flux
we have assumed a 4 keV bremsstrahlung spectrum (this is taken for simplicity;
a 1 keV disk blackbody model gives a similar result) similar to the spectral
shape of the high-intensity emission of GRS 1915+105 (Greiner, Morgan and
Remillard 1996). We then determine the normalization such that the
2--10 keV flux is identical to the flux obtained with the --4.1 powerlaw 
model (190 mCrab). The 4 keV bremsstrahlung spectrum with this normalisation
is used to model the absorbing column N$_{\rm H}$ necessary to 
reproduce the observed HRI countrate, leading to a rather small value of
N$_{\rm H}$ = 0.9$\times$10$^{22}$ cm$^{-2}$. The resulting minimum
unabsorbed flux (e.g. a 4 keV bremsstrahlung extrapolation of the
best fit PCA/RXTE spectrum below 2 keV) then follows as 
9.1$\times$10$^{-9}$ erg/cm$^2$/s (in the 0.1--10 keV range) 
corresponding to 5$\times$10$^{37}$ erg/s at 7 kpc.

\subsection{Absolute magnitude of the optical counterpart}

Very recently, the optical/IR counterpart has been identified
(Mirabel \etal 1996). 
Using the reddening conversion of Rieke \& Lubofsky (1985)
with A$_{\rm K}$ = 0.112 $\times$ A$_{\rm V}$ = 1.5$\pm$0.3
results in an extinction corrected K magnitude of 13.1 mag, and an absolute 
R, J and K magnitude (at an assumed distance of 7 kpc) of 
M$_{\rm R}$, M$_{\rm J}$ and M$_{\rm K}$ $\gax$ --4.2, --2.1 and --1.1 mag,
respectively. We caution that these numbers sensitively depend on A$_{\rm V}$,
e.g. the R--K colour reduces from --3.1 to --1.9 when using our lower
bound of A$_{\rm V}$ = 12 instead of 14. 
These blue colors are difficult to reconcile by a combination of thermal
emitters, and do not allow to constrain the companion type.
On the contrary, assuming a near-infrared color of the accretion disk
similar to those of a hot OB star (R--K=--0.8) implies E$_{\rm R-K}$ = 6.6
or A$_{\rm V}$ = 11, close to our lower bound.

\subsection{Estimates of earlier X-ray activity}

\subsubsection{Archival X-ray observations}

During the \ros\, all-sky survey the location of \grs\, was scanned
on Sep. 7/8, 1990 for a total of 330 sec. No emission is detected from \grs\,
giving an upper limit of 0.015 PSPC cts/s.

\grs\, was serendipitously in the field of view of two earlier \ros\, PSPC
observations, namely the galactic center raster pointings (PI: J. Tr\"umper) 
performed
on March 1/2, 1992 (at various off-axis angles ranging from 9\amin--50\amin) 
and a pointing (PI: C. Motch) performed on Sep. 8, 1993.
\grs\, was not detected in both observations, which have effective
exposure times at the \grs\, position of slightly above 1 ksec, thus giving 
upper limits of 5.1$\times$10$^{-3}$ and 7.7$\times$10$^{-3}$ PSPC cts/sec
for the March 1992 and Sep. 1993 observation, respectively.
This is a factor 1.2$\times$10$^4$ lower than the mean flux on March 31, 1996
(including a PSPC/HRI countrate  conversion factor of 3).

\subsubsection{The dust halo as a record of earlier X-ray activity}

Due to the longer light travel time of the dust scattered X-rays, the halo
can be used to determine earlier X-ray activity (Tr\"umper \& Sch\"onfelder 
1973) of \grs.
As can be inferred from Fig. \ref{dust}, the dust scattering halo is observed
to beyond  1000\asec\, and is clearly above the typical HRI background level
within the whole field of view.
The X-rays observed at 1000\asec\, off the nominal position of \grs\,
have a travel time which is longer than that of the unscattered emission
(assuming homogeneous distribution of dust along the line of sight) by 
$$ t_{\rm diff} \approx {d \over c} \times \left ({1 \over cos(1000\asec)} - 1 \right )
                = 98\, (d/7\,{\rm kpc})\, {\rm days}$$
Shorter timescales are reached only if the X-rays are scattered by very local 
dust. For instance, assuming scattering only within 1 kpc around the Sun, the 
delay would be $\approx$15 days (again for 7 kpc nominal distance and 
1000\asec). Given the previous notation that \grs\, went into outburst 
at the end of February 1996, the mere existence of a dust scattering halo 
extending over the whole HRI field of view is therefore surprising.
There are two alternative explanations: (1) The observed halo is exclusively 
due to scattering at local dust or (2) The X-ray outburst (or X-ray turn-on)
occurred already a few months earlier and being missed due to lack of 
observations.

If the observed scattering halo would be exclusively due to scattering
within, say, 1 kpc and no scattering near the source (i.e. around 7 kpc
from Earth) then the measured halo intensity would be a drastic
underestimate of the hypothetic steady state halo intensity. This
would imply a considerably larger distance of \grs\, which in turn would 
increase the travel time difference even more. In view of the absorbing column
determined from TTM and RXTE spectra as well as the flux ratio of the
quasi-simultaneous ROSAT and RXTE observations, this explanation seems to 
be unlikely.

Investigating the ASM/RXTE lightcurve in more detail shows \grs\, being active 
not only since
the start of the regular ASM monitoring at the end of February 1996, but also
at the $\approx$200 mCrab level during the few days of ASM operation
in January 1996. This is consistent with the second alternative that
\grs\, probably has turned on already a few months earlier.
On the other hand, there is no report on \grs\, activity during the
Sep./Oct. 1995 Granat observations of the galactic center (Sunyaev 1996).
With the above 
light travel time difference of three months relative to the ROSAT observation
on March 31, 1996 we speculate that the turn-on of \grs\, happened in
Nov./Dec. 1995.

\section{Discussion}

With our distance estimate of 6--8.5 kpc, \grs\, is certainly
another X-ray transient near the Galactic center.
Though the luminosity determination of \grs\, based on only the HRI data is
uncertain by up to a factor of 100, it is rather safe to say that \grs\,
is radiating at least near the Eddington luminosity (if not above) for
a 1 \msun\, compact object.

The finding of X-ray activity of \grs\, starting as early as at the end of 
1995 and continuing over after several months 
(given the ongoing activity as monitored with the ASM/RXTE) without
a clear trend of intensity decay argues against the classical black hole
transients which show typical decay times of the order of 30 days.
Similarly, the X-ray spectrum is unlike that of classical black hole
transients though obviously consisting of more than one component 
(Takeshima 1996). 
It is interesting to note that both the X-ray spectrum as well as the
X-ray intensity evolution are very similar to GRS 1915+105
(Harmon \etal 1994; Greiner, Morgan \& Remillard 1996), the first galactic 
source showing apparent superluminal motion (Mirabel \& Rodriguez 1994).
Given the additional similarity in the radio intensity evolution
(Hjellming \& Rupen) it seems likely that \grs\, is a good candidate
for yet another galactic superluminal motion source.

\begin{acknowledgements}
We are grateful to J. Tr\"umper for granting a \ros\, target of 
opportunity observation and to T. Takeshima for communicating the details
of the spectral fits of the PCA/RXTE data of March 31, 1996.
We thank the referee C. Motch for valuable comments.
The ASM lightcurve information is taken from quick-look results provided
by the ASM/RXTE team which we greatly acknowledge.
JG is supported by the Deutsche Agentur f\"ur
Raumfahrtangelegenheiten (DARA) GmbH under contract No. FKZ 50 OR 9201.
The ROSAT project is supported by the German Bundes\-mini\-ste\-rium f\"ur 
Bildung, Wissenschaft, For\-schung und Technologie (BMBF/DARA) and the 
Max-Planck-Society.
\end{acknowledgements}

\end{document}